# On $J/\psi$ and $\Upsilon$ transverse momentum distributions in high energy collisions

Bao-Chun Li, Ting Bai, Yuan-Yuan Guo and Fu-Hu Liu

*Department of Physics, Shanxi University, Taiyuan, Shanxi 030006, China*

Correspondence should be addressed to Bao-Chun Li; libc2010@163.com

**Abstract:** The transverse momentum distributions of final-state particles are very important for high-energy collision physics. In this work, we investigate $J/\psi$ and $\Upsilon$ meson distributions in the framework of a particle-production source, where Tsallis statistics are consistently integrated. The results are in good agreement with the experimental data of proton-proton ($p$-$p$) and proton-lead ($p$-Pb) collisions at LHC energies. The temperature of the emission source and the nonequilibrium degree of the collision system are extracted.

## 1. Introduction

The creation and study of nuclear matter at high energy densities are the purpose of Relativistic Heavy Ion Collider (RHIC) and Large Hadron Collider (LHC) [1-7]. As a new matter state, quark gluon plasma (QGP) is a thermalized system which consists of strong coupled quarks and gluons in a limited region. The suppression of J/ψ meson with respect to proton-proton ($p$-$p$) collisions is regarded as a distinctive signature of the QGP formation and brings valuable insight into properties of the nuclear matter. In proton-nucleus (*p*-A) collisions, the prompt J/ψ meson suppression has also been observed at large rapidity [8]. But, QGP is not expected to be created in the small system. The heavy quarkonium production can be suppressed by the suppression cold-nuclear-matter (CNM) effects, such as nuclear absorption, nuclear shadowing (antishadowing) and parton energy loss.

The transverse momentum $p_T$ spectra of identified particles produced in the collisions are a vital research for physicists. Now, different models have been developed to describe the $p_T$ distributions of the final-state particles in high energy collisions [9-12], such as Boltzmann distribution, Rayleigh distribution, Erlang distribution, the multisource thermal model, Tsallis statistics and soon on. Different phenomenological models of initial coherent multiple interactions and particle transport have been proposed to discuss the particle production in high-energy collisions. In condensed matter research, Tsallis statistics can deal with non-equilibrated complex systems [13]. Then, Tsallis statistics are developed to describe the particle production [14-18].

In our previous work [11], the temperature information was understood indirectly by an excitation degree. We have obtained the emission source location dependence of the exciting degree specifically. In this paper, the temperature of the emission source is given directly by combining a picture of the particle-production source with Tsallis statistics. We discuss the transverse momentum distributions of $J/\psi$ in $p$-$p$ collisions at $\sqrt{s_{NN}}$ = 7 TeV, $\sqrt{s_{NN}}$ = 8 TeV and $\sqrt{s_{NN}}$ = 13 TeV, and $p$-Pb collisions at $\sqrt{s_{NN}}$ = 5 TeV. And, the $\Upsilon$ distributions in $p$-$p$ collisions at $\sqrt{s_{NN}}$ = 8 TeV are also taken into account for comparison.

## 2. Tsallis statistics in an emission source

According to the multisource thermal model [11], at the initial stage of nucleon–nucleon (or nucleon–nucleus) collisions, a projectile cylinder and a target cylinder are formed at the rapidity $y$ space when the projectile nucleon and target nucleon pass each other. The projectile and target cylinder can be regarded as one emission source with a rapidity width. The source emits the observed particles, which follow a certain distribution.

In order to describe the transverse momentum spectra in high-energy collisions, several versions of Tsallis distribution are proposed. But, they originate from the Ref. [13], the meson number is given by

$$N = gV \int \frac{d^3 p}{(2\pi)^3} \left[ 1 + (q-1) \frac{E-\mu}{T} \right]^{1/(1-q)}, \quad (1)$$

where $g$, $V$, $p$, $E$ and $\mu$ is the degeneracy factor, the volume, the momentum, the energy and the chemical potential, respectively. $T$ and $q$ are two main parameters. $T$ is the temperature of the emission source, the entropy index $q$ is used to characterize the nonequilibrium degree. Generally, $q$ is always greater than 1 and close to 1. The corresponding momentum is

$$E \frac{d^3 N}{d^3 p} = gVE \frac{1}{(2\pi)^3} \left[ 1 + (q-1) \frac{E-\mu}{T} \right]^{1/(1-q)}. \quad (2)$$

The transverse momentum distribution is

$$\frac{d^2N}{p_T dy dp_T} = gV \frac{m_T \cosh y}{(2\pi)^2} \left[1 + (q-1)\frac{m_T \cosh y - \mu}{T}\right]^{1/(1-q)}. \quad (3)$$

When the chemical potential is neglected, at midrapidity $y = 0$, the $p_T$ distribution is

$$\frac{d^2N}{p_T dy dp_T} = \frac{gV m_T}{(2\pi)^2} \left[1 + (q-1)\frac{m_T}{T}\right]^{1/(1-q)}. \quad (4)$$

It is worth noting that the $p_T$ distribution function is only the distribution of mesons emit from an emission point $y = 0$ in the emission source, not the final-state distribution due to the nonzero rapidity width of the emission source. By summing contributions of all emission points, the transverse $p_T$ distribution is rewritten as

$$\frac{dN}{p_T dp_T} = c \int_{y_{min}}^{y_{max}} \cosh y \, dy \, m_T \left[1 + (q-1)\frac{m_T \cosh y}{T}\right]^{1/(1-q)}, \quad (5)$$

where $c = \frac{gV}{(2\pi)^2}$ is a normalize constant, $y_{max}$ and $y_{min}$ are the maximum and minimum values of the observed rapidity.

## 3. Transverse momentum spectra and discussions

Figure 1 presents the double-differential cross-section $d^2\sigma/dp_T dy$ for J/ψ mesons produced in $p$-$p$ collisions at $\sqrt{s_{NN}} = 7$ TeV. Figure 1(a), 1(b), 1(c), and 1(d) are prompt $J/\psi$ with no polarization, $J/\psi$ from $b$ with no polarization, prompt $J/\psi$ with full transverse polarization and prompt $J/\psi$ with full longitudinal polarization, respectively. The experimental data in $y$ bins are from the LHCb Collaboration [19]. The solid lines indicate our theoretical calculations. The results show an agreement with the experiment data in the observed rapidity range. The parameters $T$ and $q$ taken in the calculation are listed in Table 1. In different rapidity ranges, the values of the temperature $T$ are different and decrease with increasing the rapidity bins in all four figures of Fig. 1. The closer the emission source is to the center $y = 0$, the larger the excitation degree is. The values of $q$ are very close to 1 and do not change regularly with the rapidity bins.

Figure 2 presents the double-differential cross-section $d^2\sigma/dp_T dy$ for J/ψ mesons produced in $p$-$p$ collisions at $\sqrt{s_{NN}}$ = 8 TeV. Figure 2(a) and 2(b) are prompt $J/\psi$ and $J/\psi$ from $b$, respectively. Figure 3 presents the double-differential cross-section $B^{iS}d^2\sigma^{iS}/dp_T dy$ for ϒ mesons produced in the same collision, where $i$ = 1, 2 and 3 correspond to $\Upsilon(1S)$, $\Upsilon(2S)$, and $\Upsilon(3S)$ mesons respectively, the $B$ is the dimuon branching fraction. The experimental data are taken from Ref. [20]. The symbols and lines represent the same meanings as those in Fig. 1. The results are also in agreement with the experiment data. The parameters $T$ and $q$ taken in the calculation are listed in Table 1 and 2. The temperature $T$ also decreases with increasing the rapidity bins both for $J/\psi$ and ϒ. In the same $y$ ranges, the $T$ values are smaller for ϒ than that for $J/\psi$. For $J/\psi$, the $T$ values are generally larger than that at $\sqrt{s_{NN}}$ = 7 TeV. The values of $q$ are 1.003-1.038 for $J/\psi$ and 1.170-1.127 for ϒ.

Figure 4 presents the double-differential cross-section $d^2\sigma/dp_T dy$ for J/ψ mesons produced in $p$-$p$ collisions at $\sqrt{s_{NN}}$ =13TeV. The experimental data are taken from Ref. [21]. The symbols and lines represent the same meanings as those in Fig. 1. The results are also in agreement with the experiment data. The $T$ and $q$ values used in the calculation are listed in Table 2. The temperature $T$ decreases with increasing the rapidity bins. As the emission source draws closer to the center, the excitation degree increases. The $T$ values are larger than that at $\sqrt{s_{NN}}$ = 8 TeV in the same $y$ range. The $q$ behavior is similar to that of Fig. 1 and 2.

For comparison, Fig. 5 presents the double-differential cross-section $d^2\sigma/dp_T dy$ for J/ψ mesons produced in $p$-Pb collisions at $\sqrt{s_{NN}}$ =5TeV. The experimental data are taken from Ref. [22]. The symbols and lines represent the same meanings as those in Fig. 1. The results are also in agreement with the experiment data. The parameters $T$ and $q$ taken in the calculation are listed in Table 2. The emission source temperature $T$ decreases with increasing the rapidity bins. The values of $q$ are between 1.031 and 1.065.

## 4. Conclusions

In the framework of the emission source, where Tsallis statistics are consistently integrated, we investigate the transverse momentum spectra of $J/\Psi$ and $\Upsilon$ mesons produced in $pp$ and $p$-Pb collisions over an energy range from 5 to 13 TeV. The results agree with the experimental data of the LHCb Collaboration in LHC. By comparing with the experimental data, the emission source temperature $T$ is extracted and decreases with increasing the rapidity bins. It is consistent with the corresponding rapidity that the closer the emission source is to the $y$ center, the larger the excitation degree is. The temperature $T$ increases with increasing the collision energy. So, the excitation degree of the emission source also increases with increasing the collision energy. The parameter $q$ does not show an obvious change, which means the collision system is not very unstable.

Final-state particle production in high-energy collisions have attracted much attention, since attempt has been made to understand the properties of strongly coupled QGP by studying the production mechanisms. Thermal-statistical models have been successful in describing particle yields in various systems at different energies. The emission-source temperature is very important for understanding the matter evolution in $p$-$p$ collisions at high energy. In the rapidity space, different final-state particles emit from different positions due to stronger longitudinal flow. In our previous work [11], we have studied the transverse momentum spectra of strange particles produced in $A$-$A$ collisions at $\sqrt{s_{NN}}$ = 62.4 and 200 GeV in the improved fireball model. The temperature $T$ of the emission source was characterized indirectly by the excitation degree, which varies with location in the cylinder. In the present work, we can directly extract the specific temperature by Tsallis statistics.

In summary, the transverse momentum spectra of $J/\Psi$ and $\Upsilon$ mesons produced in $pp$ and $p$-Pb collisions at high energies have been studied in the emission-source model, where Tsallis statistics are consistently integrated. The results are compared with the experimental data of the LHCb Collaboration. The formulation is successful in the description of the distributions. At the same time, it can offer the temperature of the emission source and the information about the nonequilibrium degree in the collisions.

**Acknowledgments**

This work is supported by the National Natural Science Foundation of China under Grants No. 11247250 and No. 10975095, the National Fundamental Fund of Personnel Training under Grant No. J1103210, the Shanxi Provincial Natural Science Foundation under Grants No. 2013021006.

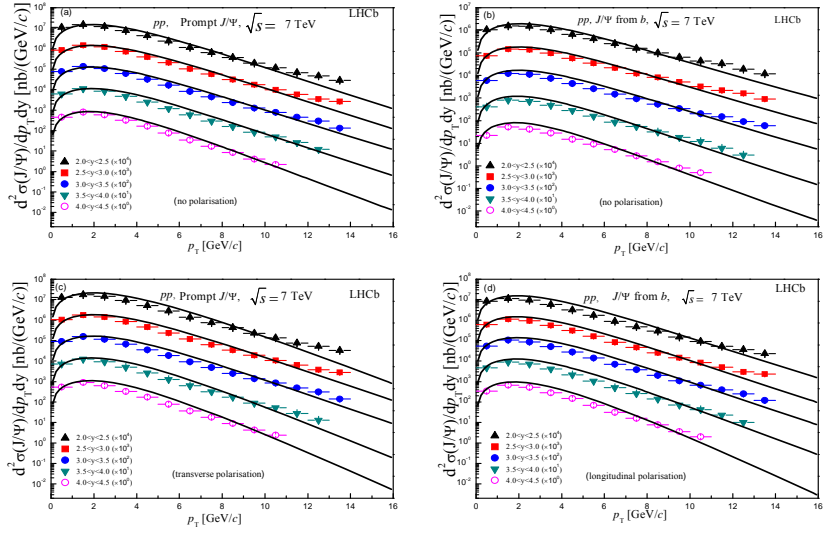

Figure 1. Double-differential cross-section $d^2\sigma/dp_T dy$ as a function of $p_T$ in bins of *y* for (a) prompt $J/\psi$ with no polarisation, (b) $J/\psi$ from *b* with no polarisation, (c) prompt $J/\psi$ with full transverse polarisation and (d) prompt $J/\psi$ with full longitudinal polarisation in *p-p* collisions at $\sqrt{s_{NN}} = 7$ TeV. The symbols indicate the experimental data of the LCHb Collaboration [19]. The solid lines indicate our theoretical calculations.

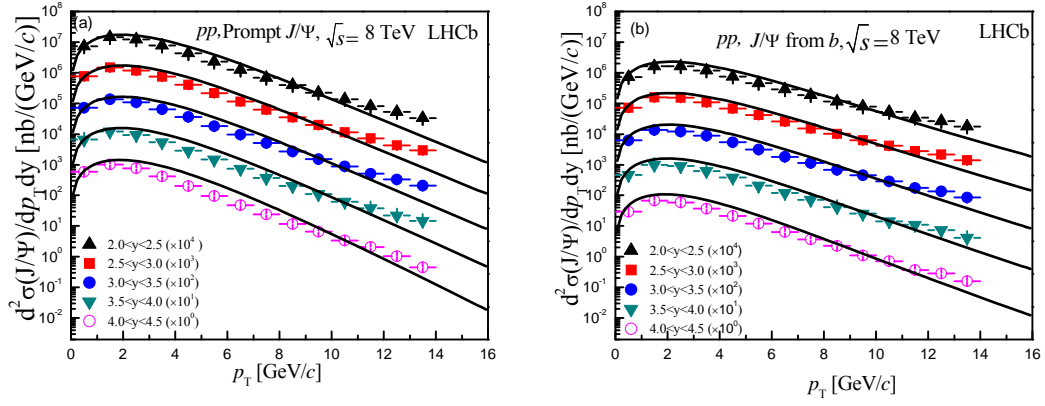

Figure 2. Double-differential cross-section $d^2\sigma/dp_T dy$ as a function of $p_T$ in bins of $y$ for (a) prompt $J/\psi$ and (b) $J/\psi$ from $b$ in $p$-$p$ collisions at $\sqrt{s_{NN}} = 8$TeV. The symbols indicate the experimental data of the LCHb Collaboration [20]. The solid lines indicate our theoretical calculations.

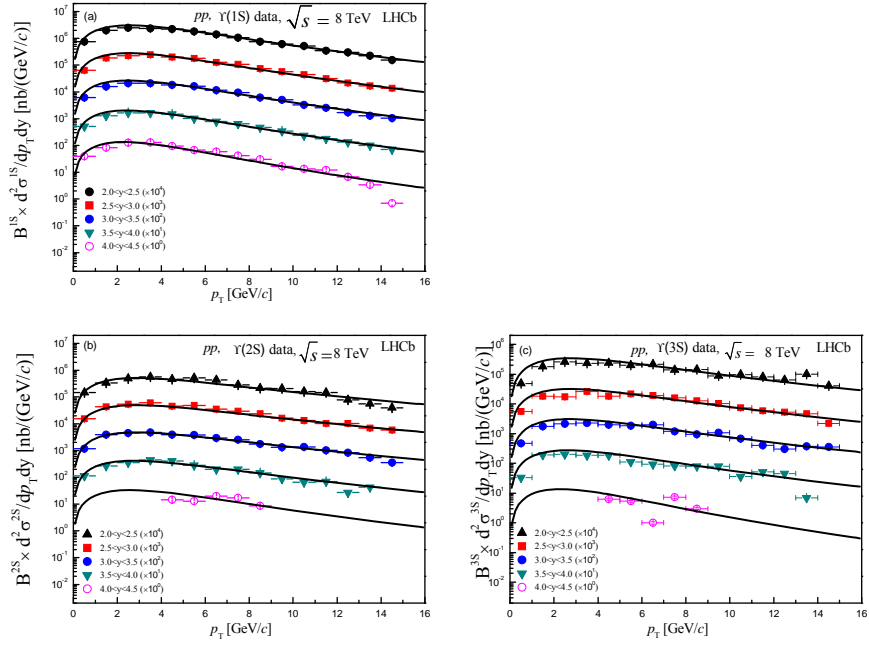

Figure 3. Double-differential cross-section $d^2\sigma/dp_T dy$ as a function of $p_T$ in bins of $y$ for (a) $\Upsilon(1S)$, (b) $\Upsilon(2S)$ and (c) $\Upsilon(3S)$ mesons in $p$-$p$ collision at $\sqrt{s_{NN}} = 8$TeV. The symbols indicate the experimental data of the LCHb Collaboration [20]. The solid lines indicate our theoretical calculations.

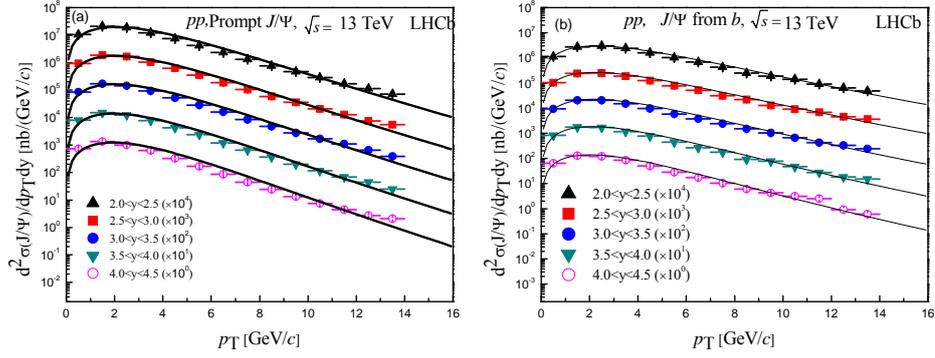

Figure 4. Same as for Fig. 2, but for $\sqrt{s_{NN}} = 13$ TeV. The experimental data are taken from the LHCb Collaboration [21].

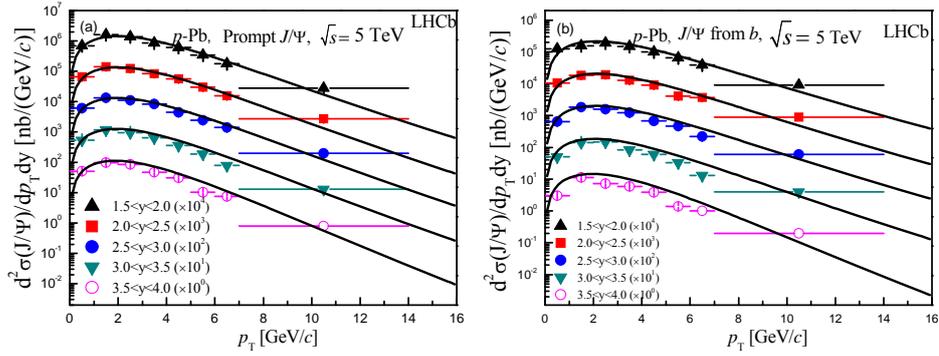

Figure 5. Same as for Fig. 2, but for $p$-Pb collisions at $\sqrt{s_{NN}} = 5$ TeV. The experimental data are taken from the LHCb Collaboration [22].

Table 1. Values of $T$ and $q$ used in Figure 1-2.

| Figure | y bins | $T$ (GeV) | $q$ |
|---|---|---|---|
| 1 (a) | 2.0 < y < 2.5 | 0.845 | 1.018 |
|  | 2.5 < y < 3.0 | 0.833 | 1.020 |
|  | 3.0 < y < 3.5 | 0.801 | 1.024 |
|  | 3.5 < y < 4.0 | 0.789 | 1.021 |
|  | 4.0 < y < 4.5 | 0.755 | 1.015 |
| 1 (b) | 2.0 < y < 2.5 | 0.863 | 1.034 |
|  | 2.5 < y < 3.0 | 0.850 | 1.033 |
|  | 3.0 < y < 3.5 | 0.838 | 1.032 |
|  | 3.5 < y < 4.0 | 0.783 | 1.028 |
|  | 4.0 < y < 4.5 | 0.725 | 1.026 |
| 1 (c) | 2.0 < y < 2.5 | 0.950 | 1.002 |
|  | 2.5 < y < 3.0 | 0.900 | 1.011 |
|  | 3.0 < y < 3.5 | 0.875 | 1.012 |
|  | 3.5 < y < 4.0 | 0.867 | 1.003 |
|  | 4.0 < y < 4.5 | 0.826 | 1.001 |
| 1 (d) | 2.0 < y < 2.5 | 0.845 | 1.020 |
|  | 2.5 < y < 3.0 | 0.840 | 1.017 |
|  | 3.0 < y < 3.5 | 0.836 | 1.016 |
|  | 3.5 < y < 4.0 | 0.829 | 1.010 |
|  | 4.0 < y < 4.5 | 0.802 | 1.001 |
| 2 (a) | 2.0 < y < 2.5 | 0.882 | 1.013 |
|  | 2.5 < y < 3.0 | 0.880 | 1.013 |
|  | 3.0 < y < 3.5 | 0.876 | 1.011 |
|  | 3.5 < y < 4.0 | 0.873 | 1.008 |
|  | 4.0 < y < 4.5 | 0.866 | 1.003 |
| 2 (b) | 2.0 < y < 2.5 | 0.885 | 1.038 |
|  | 2.5 < y < 3.0 | 0.880 | 1.036 |
|  | 3.0 < y < 3.5 | 0.874 | 1.031 |
|  | 3.5 < y < 4.0 | 0.830 | 1.030 |
|  | 4.0 < y < 4.5 | 0.805 | 1.029 |

Table 2. Values of $T$ and $q$ used in Figure 3-5.

| Figure | y bins | T | q |
|---|---|---|---|
| 3 (a) | 2.0 < y < 2.5 | 0.744 | 1.138 |
|  | 2.5 < y < 3.0 | 0.734 | 1.132 |
|  | 3.0 < y < 3.5 | 0.728 | 1.130 |
|  | 3.5 < y < 4.0 | 0.670 | 1.132 |
|  | 4.0 < y < 4.5 | 0.610 | 1.127 |
| 3 (b) | 2.0 < y < 2.5 | 0.795 | 1.170 |
|  | 2.5 < y < 3.0 | 0.790 | 1.168 |
|  | 3.0 < y < 3.5 | 0.788 | 1.165 |
|  | 3.5 < y < 4.0 | 0.784 | 1.151 |
|  | 4.0 < y < 4.5 | 0.770 | 1.133 |
| 3 (c) | 2.0 < y < 2.5 | 0.703 | 1.169 |
|  | 2.5 < y < 3.0 | 0.688 | 1.168 |
|  | 3.0 < y < 3.5 | 0.684 | 1.167 |
|  | 3.5 < y < 4.0 | 0.680 | 1.157 |
|  | 4.0 < y < 4.5 | 0.605 | 1.131 |
| 4 (a) | 2.0 < y < 2.5 | 0.862 | 1.035 |
|  | 2.5 < y < 3.0 | 0.840 | 1.034 |
|  | 3.0 < y < 3.5 | 0.828 | 1.033 |
|  | 3.5 < y < 4.0 | 0.802 | 1.032 |
|  | 4.0 < y < 4.5 | 0.782 | 1.031 |
| 4 (b) | 2.0 < y < 2.5 | 0.888 | 1.065 |
|  | 2.5 < y < 3.0 | 0.864 | 1.063 |
|  | 3.0 < y < 3.5 | 0.836 | 1.060 |
|  | 3.5 < y < 4.0 | 0.810 | 1.057 |
|  | 4.0 < y < 4.5 | 0.756 | 1.054 |
| 5 (a) | 1.5 < y < 2.0 | 0.792 | 1.040 |
|  | 2.0 < y < 2.5 | 0.786 | 1.037 |
|  | 2.5 < y < 3.0 | 0.784 | 1.036 |
|  | 3.0 < y < 3.5 | 0.782 | 1.033 |
|  | 3.5 < y < 4.0 | 0.780 | 1.025 |
| 5 (b) | 1.5 < y < 2.0 | 0.843 | 1.053 |
|  | 2.0 < y < 2.5 | 0.840 | 1.050 |
|  | 2.5 < y < 3.0 | 0.838 | 1.048 |
|  | 3.0 < y < 3.5 | 0.836 | 1.041 |
|  | 3.5 < y < 4.0 | 0.822 | 1.026 |